# The Use of Activity Trackers for Health Empowerment and Commitment: The Philippine Cycling Perspective

Ryan A. Ebardo[+]

Faculty Member, College of Computer Studies and Engineering, Jose Rizal University

Doctorate Student, College of Computer Studies, De La Salle University

**Abstract.** Activity tracking devices have found its way in the world of cycling. With its projected market demand and increasing popularity of cycling in the Philippines, cyclists are slowly adopting this technology in their daily cycling routines. Activity trackers demonstrate real-time data which allow cyclists to adjust physical efforts to achieve their personal goals. Six common features of activity trackers were formed as constructs to explore its influence on health empowerment in the context of cycling using Partial Least Squares – Structural Equation Model. A total of 393 cyclists in the Philippines participated in the study. Some features demonstrated strong evidence of positive influence in achieving health empowerment and commitment. Implications for future design and development of this technology device are discussed.

## 1. Introduction

An activity tracker belongs to one of the subsets of wearable devices. Data on physical activities is transmitted usually wirelessly from the device itself to a smart phone or a personal computer. These results are sometimes compared to another user or with members of a community to induce excitement in its usage [1]. Although developed primarily for fitness, users from sectors such as education [2], medicine [3] and sports [4] have slowly adopted activity trackers for different purposes.

In cycling, activity trackers have been utilized as early as 2007 in the form of heart rate monitors. It was widely acknowledged by cyclists, coaches and athletes as an effective tool for sports performance enhancement [5]. The recent popularity of activity trackers and its convergence with smart watches paved the way for its entry in the cycling community. In the past, cyclists monitor their cycling activities through a physically installed bike computer and are limited to data such as distance covered and speed. With the entry of smartphones, cyclists were able to incorporate features such as GPS, interactive maps and music in their cycling routines. However, research suggests negative impacts such as performance depreciation and safety risks [6] as primary concerns in the use of activity trackers in cycling. Activity trackers provide cyclists more data and independence as they have the option to change their bikes and synchronize data in a single platform. In addition, cyclists are able to monitor their fitness levels through the heart rate and their sleep quality [7].

In one of the earliest studies, transportation issues in Metro Manila primarily drove the use of bicycles [8]. The Philippines has a lower bike ownership in Asia compared to its neighbors with an estimated 23% of the population having bikes compared to 49% for the entire Asian region. The support of some local government units, the establishment of cycling lanes and the worsening traffic situation are just some of the factors that encourage people to bike in the Philippines [9].

---

[+] Corresponding author. Tel.: Tel.: + 639178019672; fax: +6328372042.
*E-mail address*: ryan.ebardo@jru.edu

A recent study explored the influence of the six common features of smart wristbands, also considered as activity trackers, to health empowerment. These characteristics are: (1) Perceived attractiveness, (2) perceived monitoring, (3) Perceived privacy protection, (4) Readability, (5) Gamification and (6) Feedback [10]. For this paper, the same characteristics are utilized and relate it to commitment using the self-regulation theory.

This study contributes to the limited literature on the use of activity trackers and its effect in achieving health empowerment and commitment among cyclists in the Philippines. Future development initiatives on activity trackers will be able to use the results of this research to add or remove features in current models of activity trackers to target the cycling community market. The model is discussed in furtherance and the methodology is explained. The findings of the survey process are followed by the conclusion and recommendations of this study.

## 2. Review of Related Literature

There is a limited research on the effects of the features of activity trackers in health empowerment and commitment in cycling activities. Most of the studies are anchored on its acceptability, outcomes and perceptions in a non-cycling environment. For example, a research published in 2016, focused on the adoption of activity trackers based on their perceived usefulness and at the same time its characteristics as a fashion accessory using the widely used Technology Adoption Model or TAM [11]. In another study, users of activity trackers were asked a set of questions on how they perceive their activity trackers and its influence on their health goals. The study also used Technology Adoption Model or TAM [12].

In terms of studying activity tracking or self-monitoring solutions among cyclists, a recent study utilized a mixed-methods approach to postulate that self-monitoring, feedback and gamification are features that cyclists deem important in their cycling activities [7]. As a persuasive technology, activity trackers provide feedback and performance monitoring data that allow users to determine their current status and compare their results with other cyclists. The gamification feature of activity trackers revolves around the concept of achieving the goals set and the improvements on status [13].

This paper hopes to contribute to the existing literature by presenting the results of a quantitative study on the influence of activity trackers on the concepts of health empowerment and commitment among cyclists in the Philippines. There is an absence of research attempting to understand the magnitude of influence in the usage of activity trackers from the cycling perspective. This paper intends to fill this gap as its contribution to research.

## 3. Research Design and Hypotheses Development

In the context of this study the common features of activity trackers are positioned as constructs that influence health empowerment and explores their effects of empowerment to two types of commitment, normative and affective. These six features are (1) Perceived attractiveness, (2) Perceived monitoring, (3) Perceived privacy protection, (4) Readability, (5) Gamification and (6) Feedback [10]. The model is adopted from a previous study and applied in the circumstance of cycling. Table 1 – Hypotheses Matrix provides a list of hypotheses to be tested.

Table 1: Hypotheses Matrix

| HYPOTHESIS | STATEMENT |
|---|---|
| H1 | Perceived attractiveness of activity trackers positively influence health empowerment among cyclists |
| H2 | Perceived monitoring features in activity trackers positively influence health empowerment among cyclists |
| H3 | Perceived privacy protection in activity trackers positively influence health empowerment among cyclists |
| H4 | Perceived readability in activity trackers positively influence health empowerment among cyclists |
| H5 | Gamification in activity trackers positively influence health empowerment among cyclists |
| H6 | Feedback mechanism features in activity trackers positively influence health empowerment among cyclists |
| H7 | Health empowerment leads to affective commitment among cyclists |
| H8 | Health empowerment leads to normative commitment among cyclists |

Empowerment is the notion of control over achieving self-made goals [10]. The concept of empowerment is anchored on a concept borrowed in psychology on self-regulation theory by Albert Bandura [14]. Recent studies have also applied self-regulation theories in sports. Self-regulation theories were instrumental in motivating athletes to achieve their goals. The experiment studied athletes and how they achieve the goals that they set through self-monitoring [15]. For this study, the concept of health empowerment can include achieving set goals such as faster speed or longer distance in cycling. Commitment is defined as sticking to one's set goals. The two types of commitment – affective and normative – is positively related to health empowerment [10]. This argument is applied in cycling and postulates that cyclists are more committed towards achieving their set goals through health empowerment. Fig. 1 – Theoretical Model summarizes the research design of this paper. The six features of activity trackers are positioned as independent constructs that positively influence health empowerment, which has a positive influence on affective commitment and normative commitment.

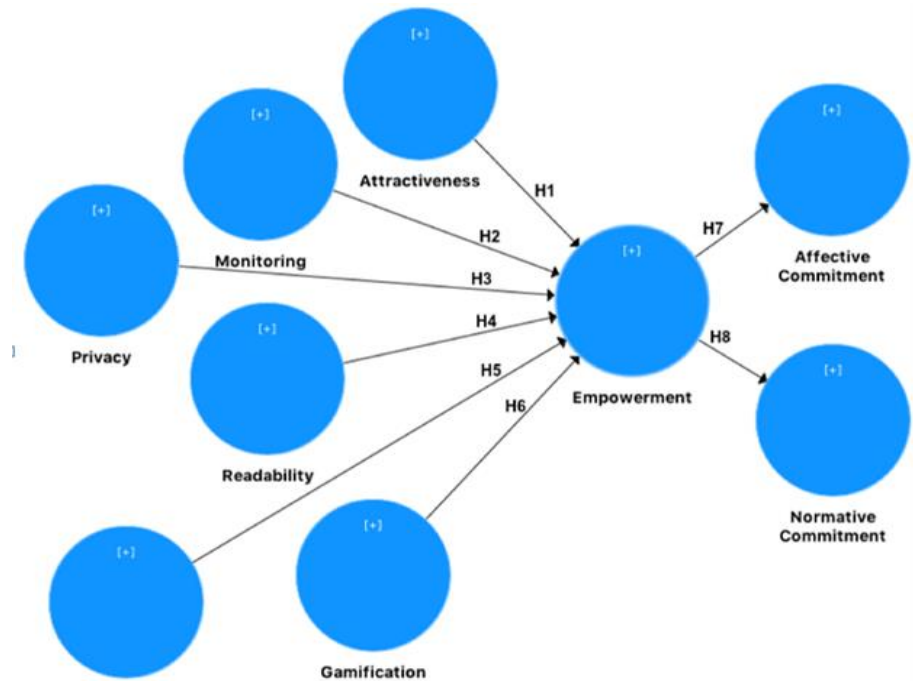

Fig. 1: Theoretical model.

## 4. Methodology

This paper approached online communities in Facebook to participate in a survey using Google forms. The participants came from Triathlon Philippines, Road Bike Philippines, Fat Bike Philippines, Pinoy-MTB, Road Bike Tiangge, and Bike Tiangge. Participants who own an activity tracking device were invited to answer the survey online. A consent section is integrated in the form to explain the purpose of the research, the requirements to answer the survey and the name of the researcher. A total of 411 responses were recorded online. After validation, 18 responses were identified as duplicate entries because they were found to have same names, but different email addresses. Responses were also checked if there are missing entries before the statistical tool was applied. A total of 393 responses were considered as valid for this research. The study adopted the instrument used by a previous study [10]. There are a total of thirty-seven (37) questions to represent the constructs in the model. Minor changes were made to fit some questions to the cycling environment. There are five (5) questions for attractiveness, four (4) questions for monitoring, four (4) questions for privacy, three (3) questions for readability, feedback (4) and four (4) questions for gamification. In addition, four (4) questions represented empowerment while affective commitment and normative commitment are represented by four (4) and five (5) questions respectively.

There are a total of 393 respondents, 251 are males and 142 are females. There are 87 participants aged 18 to 20 years old and 45 participants are 51 years old and above. In terms of the frequency, 126 indicated that they bike once a week, 64 participants ride 2 – 3 times a week, 85 bike 4 times in a week, 55 respondents bike 5 – 6 times a week and 63 respondents bike daily.

# 5. Discussion of Results

## 5.1. Instrument validation through factor analysis

To establish the validity of the instrument, a Partial Least Squares algorithm was applied to the first fifteen results using SmartPLS 3.2.6 statistical tool. Partial Least Squares– Structural Equation Model or PLS-SEM is a $2^{nd}$ generation multivariate statistical application utilized in several disciplines such as marketing and psychology. It has seen a phenomenal growth in the recent years. This statistical technique is proven to be effective in small sample size and exploratory [16].

Upon analysis of the initial results, it was found that affective commitment and privacy are below the Chronbach's Alpha critical level of 0.70 (α<0.70). To address this issue, indicators were reviewed and revealed that question 4 of the privacy construct is below 0.70 with a value of -0.105. In addition, question 4 of the affective commitment construct yielded a value of 0.166. These indicators were deleted individually with a PLS algorithm applied for each deletion. These indicators are considered as reflective and removing one will not affect the construct it represents [17]. Table 2 – Factor Analysis shows the results of factor analysis using the PLS algorithm of smartPLS to test the validity and reliability of the survey instrument.

Table 2: Factor Analysis

| Constructs | Cronbach's Alpha | Composite Reliability | Average Variance Extracted (AVE) |
|---|---|---|---|
| Attractiveness | 0.930 | 0.946 | 0.780 |
| Feedback | 0.923 | 0.945 | 0.812 |
| Gamification | 0.868 | 0.910 | 0.718 |
| Monitoring | 0.926 | 0.947 | 0.818 |
| Privacy | 0.865 | 0.917 | 0.788 |
| Readability | 0.832 | 0.899 | 0.748 |
| Empowerment | 0.882 | 0.918 | 0.738 |
| Normative Commitment | 0.897 | 0.924 | 0.709 |
| Affective Commitment | 0.783 | 0.872 | 0.696 |

## 5.2. Validation of the model using PLS-SEM

To validate the model, a bootstrapping technique was applied using smartPLS with 5,000 sample and 1-tailed test with significance level of 0.05. The statistical figures are shown in Table 3 – Model Evaluation Results. Among the features, results reveal that attractiveness (2.73), feedback (2.912) and gamification (4.402) are considered by cyclists as empowering towards achieving their personal health goals. Monitoring (1.096) is insignificant in the context of health empowerment. Health empowerment positively influence affective commitment (13.559) and normative commitment (8.416). This is consistent with a previous study [10]. In contrast, monitoring (1.096), privacy (1.173) and readability (0.343) demonstrate insignificant influence on fitness empowerment. A possible explanation is the safety risks associated with using activity trackers during cycling [6].

Table 3: Model Evaluation Results

| | | Path Coefficient | T Statistics | Significance Level | Decision |
|---|---|---|---|---|---|
| H1 | Attractiveness → Empowerment | 0.135 | 2.730 | 0.010 | Accepted |
| H2 | Monitoring → Empowerment | -0.106 | 1.096 | NS | Rejected |
| H3 | Privacy → Empowerment | 0.110 | 1.173 | NS | Rejected |
| H4 | Readability → Empowerment | 0.031 | 0.343 | NS | Rejected |
| H5 | Feedback → Empowerment | 0.140 | 2.912 | 0.010 | Accepted |
| H6 | Gamification → Empowerment | 0.497 | 4.402 | 0.010 | Accepted |
| H7 | Empowerment → Affective Commitment | 0.528 | 13.559 | 0.010 | Accepted |
| H8 | Empowerment → Normative Commitment | 0.389 | 8.416 | 0.010 | Accepted |

## 6. Conclusion and Recommendations

In conclusion, this study considered the elements of activity trackers and its influence on the concept of health empowerment. Features such as monitoring, privacy and readability are perceived as insignificant. In contrast, privacy protection, feedback and gamification positively influence health empowerment in the context of cycling. Future researches should focus on other constructs that are related to activity trackers such as ease of use [18] as well as accuracy [19] and its effect on health empowerment among cyclists. The gamification feature appears to be the construct with the strongest influence in health empowerment. The persuasive features and gamification of activity trackers improve the ride experience and encourage cyclists to perform better [13]. This study considered health empowerment and its influence to normative and affective commitment. The results confirm that given the right features, activity trackers lead to health empowerment resulting to affective commitment and normative commitment. Cyclists are most likely to commit in achieving their set goals if they feel empowered.

## 7. Acknowledgements

The author would like to acknowledge Dr. Raymund Sison of De La Salle University for his valuable critique and expertise in the development of this research paper. The researcher thanks Elizabeth Nelson of the University of Twente for the research model and instrument as well as her encouragement and insights. The author would like to express his gratitude to Jose Rizal University for the generous support in the publication of this research. The author is immensely indebted to the technical committee for the reviews and John Byron Tuazon for lending his assistance in the layout that greatly improved the manuscript. Lastly, this research will not be possible without the participation of the respondents from the different online cycling communities.